\documentclass[ aip, reprint,]{revtex4}
\usepackage{amsmath}
\usepackage{graphicx}
\usepackage{dcolumn}
\usepackage{bm}

\begin{document}

\preprint{}
\title[]{Identity for the second functional derivative of the density
functional Hartree plus exchange-correlation functional.}
\author{Daniel P. Joubert}
\email{daniel.joubert2@wits.ac.za}
\affiliation{Centre for Theoretical Physics, University of the Witwatersrand, PO Wits
2050, Johannesburg, South Africa}
\date{\today }

\begin{abstract}
It is shown that the second functional derivative of the density functional
Hartree plus exchange-correlation functional satisfies%
\begin{equation*}
\int d^{3}r^{\prime }\left( \rho _{N}\left( \mathbf{r}^{\prime }\right)
-\rho _{N-1}^{\gamma }\left( \mathbf{r}^{\prime }\right) \right) \frac{%
\delta E_{hxc}^{\gamma }[\rho _{N}]}{\delta \rho _{N}\left( \mathbf{r}%
^{\prime }\right) \delta \rho _{N}\left( \mathbf{r}\right) }=\text{constant.}
\end{equation*}%
$\rho _{N}\left( \mathbf{r}\right) $ and $\rho _{N-1}^{\gamma }\left(
\mathbf{r}\right) $ are $N$-electron and $(N-1)$-electron densities
determined from the same adiabatically scaled Hamiltonian of the interacting
electron system with $\gamma $ the scaling parameter of the
electron-electron interaction strength.
\end{abstract}

\maketitle

% Force line breaks with \\




%Use showkeys class option if keyword



%Use showkeys class option if keyword

%\volumeyear{year}
%\volumenumber{number}
%\issuenumber{number}
%\eid{identifier}
%\date[Date text]{date}
%\received[Received text]{date}

%\revised[Revised text]{date}

%\accepted[Accepted text]{date}

%\published[Published text]{date}

%\startpage{101}
%\endpage{102}
%\tableofcontents

\section{ Introduction}

The Kohn-Sham (KS) formulation \cite{KohnSham:65} of Density Functional
Theory (DFT) \cite{HohenbergKohn:64} has become the de facto tool for the
calculation of electronic structure of molecules and solids. In all
practical applications of DFT, however, approximations to the exact
functionals have to be made \cite%
{AdriennRuzsinszky2011,JohnP.Perdew2009,GaborI.Csonka2009a,Staroverov2004,TaoPerdew03}
Properties of density functionals that give an indication of the internal
structure of the functionals can give insight and help with thee development
of accurate approximations to the exact functionals \cite{JohnP.Perdew2005}.

There is little known about the properties of the second functional
derivatives of important functionals. From structural stability arguments
\cite{Levy:79,ParrYang:bk89,DreizlerGross:bk90}, for example, the second
derivative of the density functional $F^{\gamma }\left[ \rho \right] ,$ the
sum of the kinetic and interaction energy functionals, is positive definite.
Similarly for the non-interacting Kohn-Sham system, the second derivative of
the non-interacting kinetic energy functional is positive definite. In this
paper it is shown that the integral of the product of the charge density
difference of the $N$ and $(N-1)$ electron densities of the same Hamiltonian
and the second functional derivative of the Hartree plus
exchange-correlation functional of the $N$-electron density is equal to a
constant:%
\begin{equation}
\int d^{3}r^{\prime }\left( \rho _{N}\left( \mathbf{r}^{\prime }\right)
-\rho _{N-1}^{\gamma }\left( \mathbf{r}^{\prime }\right) \right) \frac{%
\delta E_{hxc}^{\gamma }[\rho _{N}]}{\delta \rho _{N}\left( \mathbf{r}%
^{\prime }\right) \delta \rho _{N}\left( \mathbf{r}\right) }=\text{constant.}
\label{ID1}
\end{equation}

Here $\rho _{N}$ and $\rho _{N-1}^{\gamma }$ are the ground state charge
densities of an interacting $N$ and $\left( N-1\right) $ electron system of
the same Hamiltonian with multiplicative external potential $v_{\text{ext}%
}^{\gamma }\left( [\rho _{N}]\right) $. The potential $v_{\text{ext}%
}^{\gamma }\left( [\rho _{N}]\right) $ is constructed to keep the charge
density of the $N$ electron system independent of the coupling strength
parameter $\gamma $ \cite%
{HarrisJones:74,LangrethPerdew:75,LangrethPerdew:77,GunnarsonLundqvist:76}
that scales the electron-electron interaction strength. At $\gamma =1$ full
strength Coulomb interaction between electrons is included and the external
potential $v_{\text{ext}}^{\gamma }\left( [\rho _{N}]\right) $ is the
external potential of the fully interacting system, while $\gamma =0$
corresponds to the non-interacting Kohn-Sham potential. \

Equation (\ref{ID1}) gives new insight into the internal structure of
density functionals and can also be used to test approximate density
functionals. The simplest test can be performed at full coupling strength, $%
\gamma =1.$ Two self-consistent calculations are required to determine $\rho
_{N}$ and $\rho _{N-1}^{1}$ and then (\ref{ID1}) can be evaluated. Ideally
an approximate exchange-correlation functional will satisfy this expression.

\section{Proof}

In the adiabatic connection approach \cite%
{HarrisJones:74,LangrethPerdew:75,LangrethPerdew:77,GunnarsonLundqvist:76}
of the constrained minimization formulation of density functional theory
\cite{HohenbergKohn:64,KohnSham:65,Levy:79,LevyPerdew:85} the Hamiltonian $%
\hat{H}^{\gamma }$ for a system of $N$ electrons is given by
\begin{equation}
\hat{H}^{\gamma }=\hat{T}+\gamma \hat{V}_{ee}+\hat{v}_{N,\text{ext}}^{\gamma
}\left[ \rho _{N}\right] .  \label{a3}
\end{equation}%
Atomic units, $\hbar =e=m=1$ are used throughout. $\hat{T}$ is the kinetic
energy operator,%
\begin{equation}
\hat{T}=-\frac{1}{2}\sum_{i=1}^{N}\nabla _{i}^{2},  \label{a4}
\end{equation}%
\ and $\gamma \hat{V}_{ee\text{ }}$is a scaled electron-electron interaction,%
\begin{equation}
\gamma \hat{V}_{ee}=\gamma \sum_{i<j}^{N}\frac{1}{\left\vert \mathbf{r}_{i}-%
\mathbf{r}_{j}\right\vert }.  \label{a2}
\end{equation}%
The the external potential
\begin{equation}
\hat{v}_{\text{ext}}^{\gamma }\left[ \rho _{N}\right] =\sum_{i=1}^{N}v_{%
\text{ext}}^{\gamma }\left( \left[ \rho _{N}\right] ;\mathbf{r}_{i}\right) ,
\label{a1}
\end{equation}%
is constructed to keep the charge density fixed at $\rho _{N}\left( \mathbf{r%
}\right) ,$ the ground state charge density of the fully interacting system (%
$\gamma =1$), for all values of the coupling constant $\gamma .$ The
external potential has the form \cite{LevyPerdew:85,GorlingLevy:93}
\begin{align}
v_{\text{ext}}^{\gamma }(\left[ \rho _{N}\right] ;\mathbf{r})& =\left(
1-\gamma \right) v_{hx}([\rho _{N}];\mathbf{r})  \notag \\
& +v_{c}^{1}([\rho _{N}];\mathbf{r)}-v_{c}^{\gamma }([\rho _{N}];\mathbf{r)+}%
v_{\text{ext}}^{1}(\left[ \rho _{N}\right] ;\mathbf{r}),  \label{e1}
\end{align}%
where $v_{\text{ext}}^{1}(\left[ \rho _{N}\right] ;\mathbf{r})=v_{\text{ext}%
}\left( \mathbf{r}\right) $ is the external potential at full coupling
strength, $\gamma =1,$ and $v_{\text{ext}}^{0}(\left[ \rho _{N}\right] ;%
\mathbf{r})$ is the non-interacting Kohn-Sham potential. The exchange plus
Hartree potential \cite{ParrYang:bk89,DreizlerGross:bk90} $v_{hx}([\rho
_{N}];\mathbf{r})$ is independent of $\gamma ,$ while the correlation
potential $v_{c}^{\gamma }([\rho _{N}];\mathbf{r)}$ depends in the scaling
parameter $\gamma .$ The chemical potential%
\begin{equation}
\mu =E_{N}^{\gamma }\left( v_{\text{ext}}^{\gamma }\left[ \rho \right]
\right) -E_{N-1}^{\gamma }\left( v_{\text{ext}}^{\gamma }\left[ \rho \right]
\right)  \label{e2}
\end{equation}%
depends on the asymptotic decay of the charge density \cite%
{ParrYang:bk89,DreizlerGross:bk90,JonesGunnarsson:89,LevyPerdewSahni:84} and
hence is independent of the coupling constant $\gamma $ \cite%
{LevyGorling:96,LevyGorlingb:96}. In Eq. (\ref{e2}) $E_{N-1}^{\gamma }$ is
the groundstate energy of the $\left( N-1\right) $-electron system with the
same single-particle external potential $v_{\text{ext}}^{\gamma }\left( %
\left[ \rho _{N}\right] ;\mathbf{r}\right) $ as the $N$-electron system:%
\begin{eqnarray}
\hat{H}_{M}^{\gamma }\left\vert \Psi _{\rho _{M}^{\gamma }}^{\gamma
}\right\rangle &=&E_{M}^{\gamma }\left\vert \Psi _{\rho _{M}^{\gamma
}}^{\gamma }\right\rangle  \notag \\
\hat{H}_{M}^{\gamma } &=&\hat{T}+\gamma \hat{V}_{\text{ee}}+\hat{v}_{M,\text{%
ext}}^{\gamma }\left[ \rho _{N}\right]  \notag \\
\hat{v}_{M,\text{ext}}^{\gamma }\left[ \rho _{N}\right] &=&\sum_{i=1}^{M}v_{%
\text{ext}}^{\gamma }\left( \left[ \rho _{N}\right] ;\mathbf{r}_{i}\right)
\label{ha}
\end{eqnarray}

The energy functional $F^{\gamma }[\rho ]$ is defined as \cite%
{Levy:79,ParrYang:bk89,DreizlerGross:bk90}
\begin{eqnarray}
F^{\gamma }[\rho ] &=&\left\langle \Psi _{\rho }^{\gamma }\left\vert \hat{T}%
+\gamma \hat{V}_{ee}\right\vert \Psi _{\rho }^{\gamma }\right\rangle  \notag
\\
&=&\min_{\Psi \rightarrow \rho }\left\langle \Psi \left\vert \hat{T}+\gamma
\hat{V}_{ee}\right\vert \Psi \right\rangle ,  \label{b1}
\end{eqnarray}%
where according to the Levy constrained minimization formulation\cite%
{Levy:79}, the wavefunction $\left\vert \Psi _{\rho }^{\gamma }\right\rangle
$ yields the density $\rho $ and minimizes $\left\langle \Psi \left\vert
\hat{T}+\gamma \hat{V}_{ee}\right\vert \Psi \right\rangle .$ For $v$%
-representable densities \cite{Levy:79,ParrYang:bk89,DreizlerGross:bk90},
the densities are derived from the groundstate eigenfunctions of the
Hamiltonian in (\ref{ha}). Note that by construction of $v_{\text{ext}%
}^{\gamma }\left( \left[ \rho _{N}\right] ;\mathbf{r}\right) ,$ Eq. (\ref{e1}%
) $\rho _{N}=\rho _{N}^{1}$ is independent of $\gamma ,$ but the groundstate
density of the $\left( N-1\right) $-electron system $\rho _{N-1}^{\gamma }$
is expected to be a function of $\gamma .$ $F^{\gamma }[\rho ]$ is usually
decomposed as \cite{ParrYang:bk89,DreizlerGross:bk90}%
\begin{equation}
F^{\gamma }[\rho ]=T^{0}\left[ \rho \right] +\gamma E_{hx}\left[ \rho \right]
+E_{c}^{\gamma }\left[ \rho \right] .  \label{b2}
\end{equation}%
The correlation energy $E_{c}^{\gamma }\left[ \rho \right] $ is defined as
\cite{LevyPerdew:85}
\begin{eqnarray}
E_{c}^{\gamma }\left[ \rho \right] &=&\left\langle \Psi _{\rho }^{\gamma
}\left\vert \hat{T}+\gamma \hat{V}_{ee}\right\vert \Psi _{\rho }^{\gamma
}\right\rangle  \notag \\
&&-\left\langle \Psi _{\rho }^{0}\left\vert \hat{T}+\gamma \hat{V}%
_{ee}\right\vert \Psi _{\rho }^{0}\right\rangle ,  \label{ec1}
\end{eqnarray}%
where $\left\vert \Psi _{\rho }^{0}\right\rangle $ is the Kohn-Sham
independent particle groundstate wavefunction that yields the same density
as the interacting system at coupling strength $\gamma .$ $E_{hx}\left[ \rho %
\right] $ is the sum of the Hartree and exchange energy%
\begin{equation}
E_{hx}\left[ \rho \right] =\left\langle \Psi _{\rho }^{0}\left\vert \hat{V}%
_{ee}\right\vert \Psi _{\rho }^{0}\right\rangle  \label{b3}
\end{equation}%
and the kinetic energy functional $T^{0}\left[ \rho \right] $ is given by%
\begin{equation}
T^{0}\left[ \rho \right] =\left\langle \Psi _{\rho }^{0}\left\vert \hat{T}%
\right\vert \Psi _{\rho }^{0}\right\rangle .  \label{b4}
\end{equation}%
The full kinetic energy
\begin{eqnarray}
T^{\gamma }\left[ \rho \right] &=&\left\langle \Psi _{\rho }^{\gamma
}\left\vert \hat{T}\right\vert \Psi _{\rho }^{\gamma }\right\rangle  \notag
\\
&=&T^{0}\left[ \rho \right] +T_{c}^{\gamma }\left[ \rho \right] ,  \label{b5}
\end{eqnarray}%
with the correlation part of the kinetic energy defined as%
\begin{equation}
T_{c}^{\gamma }\left[ \rho \right] =\left\langle \Psi _{\rho }^{\gamma
}\left\vert \hat{T}\right\vert \Psi _{\rho }^{\gamma }\right\rangle
-\left\langle \Psi _{\rho }^{0}\left\vert \hat{T}\right\vert \Psi _{\rho
}^{0}\right\rangle .  \label{ec2}
\end{equation}

Assuming that $F^{\gamma }[\rho ]$ is defined for non-integer electrons \cite%
{PPLB:82,ParrYang:bk89,DreizlerGross:bk90}, at the solution point%
\begin{equation}
\frac{\delta F^{\gamma }\left[ \rho \right] }{\delta \rho \left( \mathbf{r}%
\right) }+v_{\text{ext}}^{\gamma }\left( \left[ \rho \right] ;\mathbf{r}%
\right) =\mu  \label{b6}
\end{equation}%
where $\mu $ is the chemical potential. Note that by definition of $%
F^{\gamma }[\rho ]$%
\begin{eqnarray}
E_{N}^{\gamma }\left( v_{\text{ext}}^{\gamma }\left[ \rho _{N}\right]
\right) &=&F^{\gamma }[\rho _{N}]+\int d^{3}r\rho _{N}\left( \mathbf{r}%
\right) v_{\text{ext}}^{\gamma }\left( \left[ \rho _{N}\right] ;\mathbf{r}%
\right)  \notag \\
E_{N-1}^{\gamma }\left( v_{\text{ext}}^{\gamma }\left[ \rho _{N}\right]
\right) &=&F^{\gamma }[\rho _{N-1}^{\gamma }]+\int d^{3}r\rho _{N-1}^{\gamma
}\left( \mathbf{r}\right) v_{\text{ext}}^{\gamma }\left( \left[ \rho _{N}%
\right] ;\mathbf{r}\right) .  \label{b7}
\end{eqnarray}%
From (\ref{e2}) and (\ref{b7}),%
\begin{equation}
F^{\gamma }[\rho _{N}]-F^{\gamma }[\rho _{N-1}^{\gamma }]=\mu -\int
d^{3}r\left( \rho _{N}\left( \mathbf{r}\right) -\rho _{N-1}^{\gamma }\left(
\mathbf{r}\right) \right) v_{\text{ext}}^{\gamma }\left( \left[ \rho _{N}%
\right] ;\mathbf{r}\right) .  \label{b8}
\end{equation}%
Since%
\begin{equation}
\int d^{3}r\left( \rho _{N}\left( \mathbf{r}\right) -\rho _{N-1}^{\gamma
}\left( \mathbf{r}\right) \right) =1,  \label{b9}
\end{equation}%
it follows from (\ref{b6}), (\ref{b8}) and (\ref{b9}) that%
\begin{equation}
F^{\gamma }[\rho _{N}]-F^{\gamma }[\rho _{N-1}^{\gamma }]=\int d^{3}r\left(
\rho _{N}\left( \mathbf{r}\right) -\rho _{N-1}^{\gamma }\left( \mathbf{r}%
\right) \right) \frac{\delta F^{\gamma }\left[ \rho _{N}\right] }{\delta
\rho _{N}\left( \mathbf{r}\right) }.  \label{b10}
\end{equation}

In a recent paper \cite{Joubert2011a} the author showed that
\begin{eqnarray}
&&V_{ee}^{\gamma }\left[ \rho _{N}\right] -V_{ee}^{\gamma }\left[ \rho
_{N-1}^{\gamma }\right]  \notag \\
&=&\int d^{3}r\frac{\delta V_{ee}^{\gamma }\left[ \rho _{N}\right] }{\delta
\rho _{N}^{1}\left( \mathbf{r}\right) }\left( \rho _{N}\left( \mathbf{r}%
\right) -\rho _{N-1}^{\gamma }\left( \mathbf{r}\right) \right) .  \label{vee}
\end{eqnarray}%
The proof of this equation is based on the virial theorem \cite%
{LevyPerdew:85} of the interacting system.

The charge density $\rho _{N}$ is independent of $\gamma $ by construction,
therefore from (\ref{b2}) and (\ref{b3})%
\begin{equation}
\frac{\partial }{\partial \gamma }F^{\gamma }[\rho _{N}]=E_{hx}[\rho _{N}]+%
\frac{\partial }{\partial \gamma }E_{c}^{\gamma }[\rho _{N}].  \label{b11}
\end{equation}%
The $\left( N-1\right) $-electron density $\rho _{N-1}^{\gamma }$ is a
function of $\gamma ,$ hence it follows from (\ref{b2}) and (\ref{b3}) that%
\begin{eqnarray}
\frac{\partial }{\partial \gamma }F^{\gamma }[\rho _{N-1}^{\gamma }] &=&\int
d^{3}r\frac{\partial \rho _{N-1}^{\gamma }\left( \mathbf{r}\right) }{%
\partial \gamma }\frac{\delta T^{0}\left[ \rho _{N-1}^{\gamma }\right] }{%
\delta \rho _{N-1}^{\gamma }\left( \mathbf{r}\right) }+E_{hx}[\rho
_{N-1}^{\gamma }]  \notag \\
&&+\gamma \int d^{3}r\frac{\partial \rho _{N-1}^{\gamma }\left( \mathbf{r}%
\right) }{\partial \gamma }v_{hx}\left( \left[ \rho _{N-1}^{\gamma }\right] ;%
\mathbf{r}\right) +\frac{\partial }{\partial \gamma }E_{c}^{\gamma }[\rho
_{N-1}^{\gamma }],  \label{b12}
\end{eqnarray}%
where $v_{hx}\left( \left[ \rho \right] ;\mathbf{r}\right) =\frac{\delta
E_{hx}\left[ \rho \right] }{\delta \rho \left( \mathbf{r}\right) }.$ It
follows from the definition of $E_{c}^{\gamma }[\rho _{N}],$ Eq. (\ref{ec1})
that \cite{LevyPerdew:85}
\begin{equation}
\frac{\partial }{\partial \gamma }E_{c}^{\gamma }[\rho _{N}]=\frac{1}{\gamma
}\left( E_{c}^{\gamma }[\rho _{N}]-T_{c}^{\gamma }[\rho _{N}]\right) .
\label{b13}
\end{equation}%
The derivative $\frac{\partial }{\partial \gamma }E_{c}^{\gamma }[\rho
_{N-1}^{\gamma }]$ is less simple since $\rho _{N-1}^{\gamma }$ is a
function of $\gamma $. From (\ref{ec1}) it is clear that $E_{c}^{\gamma
}[\rho _{N-1}^{\gamma }]$ depends on $\gamma $ as the scaling parameter of
the Coulomb interaction strength and via the dependence of the wavefunction $%
\left\vert \Psi _{\rho _{N-1}^{\gamma }}^{\gamma }\right\rangle $ on $\gamma
.$ It can be shown that\cite{Joubert2011}

\begin{eqnarray}
&&\frac{\partial }{\partial \gamma }E_{c}^{\gamma }\left[ \rho
_{N-1}^{\gamma }\right]  \notag \\
&=&\frac{E_{c}^{\gamma }\left[ \rho _{N-1}^{\gamma }\right] -T_{c}^{\gamma }%
\left[ \rho _{N-1}^{\gamma }\right] }{\gamma }  \notag \\
&&+\int d^{3}r\frac{\partial \rho _{N-1}^{\gamma }\left( \mathbf{r}\right) }{%
\partial \gamma }\left( v_{c}^{\gamma }\left( \left[ \rho _{N}\right] ;%
\mathbf{r}\right) +\gamma v_{hx}\left( \left[ \rho _{N}\right] ;\mathbf{r}%
\right) -\gamma v_{hx}\left( \left[ \rho _{N-1}^{\gamma }\right] ;\mathbf{r}%
\right) \right)  \label{b13a}
\end{eqnarray}

From (\ref{b11}), (\ref{b12}), (\ref{b2}), (\ref{b13}) and (\ref{b13a}) it
follows that

\begin{eqnarray}
&&E_{hxc}^{\gamma }[\rho _{N}]-T_{c}^{\gamma }[\rho _{N}]-\left(
E_{hxc}^{\gamma }[\rho _{N-1}^{\gamma }]-T_{c}^{\gamma }\left[ \rho
_{N-1}^{\gamma }\right] \right)  \notag \\
&=&\int d^{3}r\left( \rho _{N}\left( \mathbf{r}\right) -\rho _{N-1}^{\gamma
}\left( \mathbf{r}\right) \right) \frac{\delta }{\delta \rho _{N}\left(
\mathbf{r}\right) }\left( E_{hxc}^{\gamma }[\rho _{N}]-T_{c}^{\gamma }[\rho
_{N}]\right)  \notag \\
&&+\gamma \int d^{3}r\frac{\partial \rho _{N-1}^{\gamma }\left( \mathbf{r}%
\right) }{\partial \gamma }\left( \frac{\delta T^{0}\left[ \rho
_{N-1}^{\gamma }\right] }{\delta \rho _{N-1}^{\gamma }\left( \mathbf{r}%
\right) }-\frac{\delta T^{0}\left[ \rho _{N}\right] }{\delta \rho _{N}\left(
\mathbf{r}\right) }\right) .  \label{b14}
\end{eqnarray}%
Here $E_{hxc}^{\gamma }[\rho _{N}]=\gamma E_{hx}[\rho ]+E_{c}^{\gamma }[\rho
].$ From (\ref{b1}) and (\ref{b2})%
\begin{equation}
\gamma V_{ee}^{\gamma }\left[ \rho \right] =E_{hxc}^{\gamma }[\rho
_{N}]-T_{c}^{\gamma }[\rho ],  \label{b15}
\end{equation}%
and therefore from (\ref{b14})\ and (\ref{vee})

\begin{equation}
\int d^{3}r\frac{\partial \rho _{N-1}^{\gamma }\left( \mathbf{r}\right) }{%
\partial \gamma }\left( \frac{\delta T^{0}\left[ \rho _{N}\right] }{\delta
\rho _{N}\left( \mathbf{r}\right) }-\frac{\delta T^{0}\left[ \rho
_{N-1}^{\gamma }\right] }{\delta \rho _{N-1}^{\gamma }\left( \mathbf{r}%
\right) }\right) =0.  \label{p12}
\end{equation}%
By construction, from (\ref{b6}) and (\ref{b2})%
\begin{eqnarray}
\frac{\delta T^{0}\left[ \rho _{N}\right] }{\delta \rho _{N}\left( \mathbf{r}%
\right) }+v_{hxc}^{\gamma }\left( \left[ \rho _{N}\right] ;\mathbf{r}\right)
+v_{\text{ext}}^{\gamma }\left( \left[ \rho _{N}\right] ;\mathbf{r}\right)
&=&\mu _{N}^{\gamma }  \notag \\
\frac{\delta T^{0}\left[ \rho _{N-1}^{\gamma }\right] }{\delta \rho
_{N-1}^{\gamma }\left( \mathbf{r}\right) }+v_{hxc}^{\gamma }\left( \left[
\rho _{N-1}^{\gamma }\right] ;\mathbf{r}\right) +v_{\text{ext}}^{\gamma
}\left( \left[ \rho _{N}\right] ;\mathbf{r}\right) &=&\mu _{N-1}^{\gamma }
\label{p13}
\end{eqnarray}%
where $v_{hxc}^{\gamma }\left( \left[ \rho \right] ;\mathbf{r}\right) =\frac{%
\delta E_{hxc}^{\gamma }\left[ \rho \right] }{\delta \rho \left( \mathbf{r}%
\right) }$ and hence%
\begin{eqnarray}
&&\frac{\delta T^{0}\left[ \rho _{N}\right] }{\delta \rho _{N}\left( \mathbf{%
r}\right) }-\frac{\delta T^{0}\left[ \rho _{N-1}^{\gamma }\right] }{\delta
\rho _{N-1}^{\gamma }\left( \mathbf{r}\right) }  \notag \\
&=&v_{hxc}^{\gamma }\left( \left[ \rho _{N-1}^{\gamma }\right] ;\mathbf{r}%
\right) -v_{hxc}^{\gamma }\left( \left[ \rho _{N}\right] ;\mathbf{r}\right)
+\mu _{N}-\mu _{N-1}^{\gamma }.  \label{p14}
\end{eqnarray}%
The dependence of $\rho _{N-1}^{\gamma }\left( \mathbf{r}\right) $ on $%
\gamma $ is through the $\gamma $ dependence of the potential $v_{\text{ext}%
}^{\gamma }\left( \left[ \rho _{N}^{\gamma }\right] ;\mathbf{r}\right)
=v^{\gamma }\left( \mathbf{r}\right) .$ Using (\ref{p14}) in (\ref{p12}) and
since%
\begin{equation}
\frac{\partial }{\partial \gamma }\left( N-1\right) =\int d^{3}r\frac{%
\partial \rho _{N-1}^{\gamma }\left( \mathbf{r}\right) }{\partial \gamma }=0,
\label{p15}
\end{equation}%
it follows
\begin{equation}
\int d^{3}r^{\prime }\int d^{3}r\left( v_{hxc}^{\gamma }\left( \left[ \rho
_{N}\right] ;\mathbf{r}\right) -v_{hxc}^{\gamma }\left( \left[ \rho
_{N-1}^{\gamma }\right] ;\mathbf{r}\right) \right) \left. \frac{\delta \rho
_{N-1}^{\gamma }\left( \mathbf{r}\right) }{\delta v^{\gamma }\left( \mathbf{r%
}^{\prime }\right) }\right\vert _{N-1}\frac{\partial v^{\gamma }\left(
\mathbf{r}^{\prime }\right) }{\partial \gamma }=0.  \label{p16}
\end{equation}%
The functional derivative with respect to $v^{\gamma }$ is taken at constant
particle number since $\rho _{N-1}^{\gamma }$ and $\rho _{N}$ are derived
from the groundstates of the Hamiltonian in Eq. (\ref{ha}).

The author recently showed that \cite{Joubert2011b}
\begin{eqnarray}
&&E_{hxc}^{\gamma }[\rho _{N}]-E_{hxc}^{\gamma }[\rho _{N-1}^{\gamma }]
\notag \\
&=&\int d^{3}r\left( \rho _{N}\left( \mathbf{r}\right) -\rho _{N-1}^{\gamma
}\left( \mathbf{r}\right) \right) \frac{\delta }{\delta \rho _{N}\left(
\mathbf{r}\right) }E_{hxc}^{\gamma }[\rho _{N}].  \label{v8}
\end{eqnarray}%
Take the functional derivative of (\ref{v8}) with respect to $v^{\gamma
}\left( \mathbf{r}\right) $ and use he functional chain rule:%
\begin{eqnarray}
&&\int d^{3}r^{\prime }\left( v_{hxc}^{\gamma }\left( \left[ \rho _{N}\right]
;\mathbf{r}\right) -v_{hxc}^{\gamma }\left( \left[ \rho _{N-1}^{\gamma }%
\right] ;\mathbf{r}\right) \right) \left. \frac{\delta \rho _{N-1}^{\gamma
}\left( \mathbf{r}^{\prime }\right) }{\delta v^{\gamma }\left( \mathbf{r}%
\right) }\right\vert _{N-1}  \notag \\
&=&\int d^{3}r^{\prime \prime }\int d^{3}r^{\prime }\left( \rho _{N}\left(
\mathbf{r}^{\prime }\right) -\rho _{N-1}^{\gamma }\left( \mathbf{r}^{\prime
}\right) \right) \frac{\delta v_{hxc}^{\gamma }[\rho _{N};\mathbf{r}^{\prime
}]}{\delta \rho _{N}\left( \mathbf{r}^{\prime \prime }\right) }\left. \frac{%
\delta \rho _{N}\left( \mathbf{r}^{\prime \prime }\right) }{\delta v^{\gamma
}\left( \mathbf{r}\right) }\right\vert _{N}.  \label{t2}
\end{eqnarray}%
Multiply both sides by $\frac{\partial v^{\gamma }\left( \mathbf{r}^{\prime
\prime }\right) }{\partial \gamma }$ and integrate over $\mathbf{r}^{\prime
\prime }.$ It follows from (\ref{t2}) and (\ref{t6}) that
\begin{equation}
\int d^{3}r\int d^{3}r^{\prime }\int d^{3}r^{\prime \prime }\left( \rho
_{N}\left( \mathbf{r}^{\prime }\right) -\rho _{N-1}^{\gamma }\left( \mathbf{r%
}^{\prime }\right) \right) \frac{\delta v_{hxc}^{\gamma }[\rho _{N};\mathbf{r%
}^{\prime }]}{\delta \rho _{N}\left( \mathbf{r}\right) }\chi _{N}^{\gamma
}\left( \mathbf{r},\mathbf{r}^{\prime \prime }\right) \frac{\partial
v^{\gamma }\left( \mathbf{r}^{\prime \prime }\right) }{\partial \gamma }=0
\label{t5}
\end{equation}%
where%
\begin{equation}
\chi _{N}^{\gamma }\left( \mathbf{r},\mathbf{r}^{\prime }\right) =\left.
\frac{\delta \rho _{N}\left( \mathbf{r}\right) }{\delta v_{\text{ext}%
}^{\gamma }\left( \left[ \rho _{N}\right] ;\mathbf{r}^{\prime }\right) }%
\right\vert _{N}  \label{drs1}
\end{equation}%
is the density response function of the $N$-electron system. From stability
considerations $\chi _{N}^{\gamma }\left( \mathbf{r},\mathbf{r}^{\prime
}\right) $ is negative semi-definite and has one zero eigenvalue which
corresponds to the invariance of the density when the potential is changed
by a constant \cite{ParrYang:bk89,DreizlerGross:bk90}. However, from (\ref%
{e1})
\begin{equation}
\frac{\partial v^{\gamma }\left( \mathbf{r}^{\prime \prime }\right) }{%
\partial \gamma }=-v_{hx}([\rho _{N}];\mathbf{r})-\frac{\partial }{\partial
\gamma }v_{c}^{\gamma }([\rho _{N}];\mathbf{r)\neq }0,  \label{t6}
\end{equation}%
Hence, from the properties of $\chi _{N}^{\gamma }\left( \mathbf{r},\mathbf{r%
}^{\prime }\right) $ Eq. (\ref{t5}) implies the main result of this paper,%
\begin{equation}
\int d^{3}r^{\prime }\left( \rho _{N}\left( \mathbf{r}^{\prime }\right)
-\rho _{N-1}^{\gamma }\left( \mathbf{r}^{\prime }\right) \right) \frac{%
\delta v_{hxc}^{\gamma }[\rho _{N};\mathbf{r}^{\prime }]}{\delta \rho
_{N}\left( \mathbf{r}\right) }=\text{constant}  \label{t7}
\end{equation}%
since $\frac{\delta v_{hxc}^{\gamma }[\rho _{N};\mathbf{r}^{\prime }]}{%
\delta \rho _{N}\left( \mathbf{r}\right) }=\frac{\delta E_{hxc}^{\gamma
}[\rho _{N}]}{\delta \rho _{N}\left( \mathbf{r}^{\prime }\right) \delta \rho
_{N}\left( \mathbf{r}\right) }.$

\section{Discussion and summary}

In the derivation of Eq. (\ref{t7}) use was made of the groundstate
wavefunctions of the same Hamiltonian, for example in the derivation of (\ref%
{b13a}). Accordingly the relation (\ref{t7}) has been proven for $v$%
-representable pure state densities \cite{ParrYang:bk89,DreizlerGross:bk90}
only. Whether it is valid for non-$v$-representable densities and
non-integer densities remains an open question which is under investigation.
Throughout the assumption was made that all functional derivatives are well
behaved and this implies that the functionals are defined for non-integer
particle numbers as discussed in \cite%
{PPLB:82,ParrYang:bk89,DreizlerGross:bk90}.

In summary, an integral expression was derived for the second derivative of
the pure state density functional Hartree plus exchange-correlation energy
which provides new insight into the structure of density functionals. It
also places constraints on potential approximate exchange-correlation
functionals and can be used to test existing and potential approximate
functionals.

%\bibliographystyle{apsrev}
%\bibliography{dft}

\end{document}